\documentclass[preprint,aps]{revtex4}

\newcommand{\be}{\begin{equation}} \newcommand{\ee}{\end{equation}} 
\newcommand{\bea}{\begin{eqnarray}}\newcommand{\eea}{\end{eqnarray}}

\begin{document}
\preprint{SINP/TNP/03-33, hep-th/0309183}
\title{ Super-Calogero model with $OSp(2|2)$ supersymmetry : is the
construction unique?}
\author{ Pijush K. Ghosh}
\email{pijush@theory.saha.ernet.in}
\affiliation{Theory Division, Saha Institute of Nuclear Physics,\\ 
Kolkata 700 064, India.}
\begin{abstract} 
We show that the construction of super-Calogero model with $OSp(2|2)$
supersymmetry is not unique. In particular,
we find a new co-ordinate representation of the generators of the
$OSp(2|2)$ superalgebra that appears as the dynamical supersymmetry of the
rational super-Calogero model. Both the quadratic and the cubic Casimir
operators of the $OSp(2|2)$ are necessarily zero in this new representation,
while they are, in general, nonzero for the super-Calogero model that is
currently studied in the literature.
The Scasimir operator that exists in the new co-ordinate
representation is not present in the case of the existing super-Calogero model.
We also discuss the case of $N$ free superoscillators and 
superconformal quantum mechanics for which the same conclusions are valid.
\end{abstract}
\maketitle
\newpage

\section{Introduction}

Ever since the supersymmetric extension of the rational Calogero
model\cite{cs,cs1,pr,poly,sw,sw1,bbm} was introduced in \cite{fm}, various
aspects of this class of supermodels have been
studied\cite{bh,susy,turb,hs,me,sasa,other,gib,pkg,nw,sc,cana,degu,rus}.
These supermodels are exactly
solvable\cite{fm,bh,susy,turb} and play an important role in our understanding
of the integrable structure of many-particle quantum mechanics\cite{hs,me,sasa}.
Further, they are relevant in the study of the superstrings\cite{other},
black-holes\cite{gib}, many-particle superconformal quantum
mechanics\cite{pkg,nw,sc}, superpolynomials\cite{cana}, spin
chains\cite{degu,rus} etc.

The rational super-Calogero model has a dynamical $OSp(2|2)$
supersymmetry\cite{fm}. The super-Hamiltonian and the other generators
satisfying the structure equations of $OSp(2|2)$\cite{superlie,rev} are
represented in terms of a set of bosonic and fermionic variables. In all
subsequent discussions of the super-Calogero model introduced in \cite{fm},
the same co-ordinate representation of the $OSp(2|2)$ generators have been used.
Although the construction of \cite{fm} is a very standard one, it is pertinent
to ask whether the $OSp(2|2)$ superextension of the rational Calogero model is
unique or not? In other words, is it possible to have a different co-ordinate
representation of the generators of the $OSp(2|2)$ than the one originally
introduced in \cite{fm}? We indeed find that such a possibility exists and the
$OSp(2|2)$ superextension of the Calogero model is not unique.

In particular, we show that the generators of the $OSp(2|2)$ algebra can be
represented in terms of the co-ordinates of the super-Calogero model in two
different ways. One of them can be shown to be equivalent to
the construction of Ref. \cite{fm} and we refer to this particular co-ordinate
representation of the generators as `standard superextension' in all subsequent
references. The corresponding super-Hamiltonian will be referred as
standard super-Calogero model. The other representation of the
generators, although appeared in the study of
the supersymmetric magnetic monopoles\cite{dhowker}, 
has never been discussed in the literature in the context of general
many-particle supersymmetric quantum mechanics or Calogero models.
Both the quadratic and the cubic Casimir operators in this case 
are necessarily zero. Thus, the representation of the
group $OSp(2|2)$ is necessarily `atypical'\cite{superlie,rev}, i.e.,
the Casimir operators do not specify the spectrum completely.
We refer to this case as `atypical superextension' for all future purposes.
We refer the corresponding super-Hamiltonian as the `chiral super-Calogero
Hamiltonian', since it is known from the study of supersymmetric magnetic
monopoles\cite{dhowker} that the eigenstates for such a representation have
definite chirality.

The differences between these two co-ordinate representations are the following.
First of all, the Hamiltonian of the standard super-Calogero
model and the chiral super-Calogero model are different. Further, in contrast
to the `atypical superextension', the quadratic
and the cubic Casimir operators of the `standard superextension' in terms
of the co-ordinate representation are not necessarily zero. The representation
of the algebra can be either `typical' or `atypical' depending on the
particular condition used on the wavefunction. The eigenvalues of
the Casimir's can be zero only for the special case corresponding
to the `atypical' representation of the group $OSp(2|2)$.
Moreover, for the `atypical superextension', there exists a Scasimir
operator\cite{scasi}. This is an even operator which
commutes with all the bosonic generators and anti-commutes with all the
fermionic generators of the $OSp(2|2)$. This even operator is called Scasimir,
because it squares to the Casimir of the $OSp(1|1)$, a subgroup
of $OSp(2|2)$. The Scasimir does not exists for the $OSp(2|2)$ in the case of
the standard superextension.

We also show that the angular part of the super-Calogero model, in the
N-dimensional hyper-spherical coordinate, can be factorized in terms of
first order differential operators that commute with the total Hamiltonian.
Thus, eigen spectrum of these first order operators can be used to find the
eigen spectrum of the angular part of the super-Calogero model.The result is
valid for both the standard and the atypical superextension with the difference
that the factorization can be done in two independent ways for the former
case, while only one factorization is possible for the later case.
To the best of our knowledge, this has never been realized before for the
standard superextension. The reformulation of the problem in terms of two
real supercharges made us to see this subtle connection. We also show
that super-Calogero model, for both the standard and the atypical
superextension, can be deformed suitably to admit ${\cal{N}}=0$
supersymmetry\cite{spec}.

Our plan of presenting the results is the following. We briefly review
about the $OSp(2|2)$ supersymmetry, the Casimir and the Scasimir operators in
the next section. In section III, we explicitly write down the coordinate
representations of the $OSp(2|2)$ generators and present our results.
The standard and the atypical superextension are discussed in sections
III.A and III.B, respectively. The results presented till the end of the section
III.B are quite general. We consider the specific cases of superoscillators,
superconformal quantum mechanics\cite{dff,fr} and super-Calogero models
in section IV. Finally, we conclude with discussions in section V.

\section{Susy algebra}

The $OSp(2|2)$ algebra is described in terms of the set of fermionic generators
$f \equiv \{Q_1, Q_2, S_1, S_2 \}$ and the set of bosonic generators
$b \equiv \{ H, D, K, Y \}$. The
structure-equations of the $OSp(2|2)$ are\cite{superlie,rev},
\bea
&& \{Q_{\alpha}, Q_{\beta} \} = 2 \delta_{\alpha \beta} H, \ \
\{S_{\alpha}, S_{\beta}\} = 2 \delta_{\alpha \beta} K, \ \
\{Q_{\alpha},S_{\beta}\} = - 2 \delta_{\alpha \beta} D + 2 \epsilon_{\alpha
\beta} Y,\nonumber \\
&& [H, Q_{\alpha}]=0, \ [H,S_{\alpha}]=-i Q_{\alpha}, \
[K,Q_{\alpha}] = i S_{\alpha}, \ [K,S_{\alpha}]=0,\nonumber \\
&& [D, Q_{\alpha}] = - \frac{i}{2} Q_{\alpha}, \
[D, S_{\alpha}] = \frac{i}{2} S_{\alpha}, \
[Y,Q_{\alpha}] = \frac{i}{2} \epsilon_{\alpha \beta}
Q_{\beta}, \ \ [Y, S_{\alpha}]=\frac{i}{2} \epsilon_{\alpha \beta}
S_{\beta},\nonumber \\
&& [Y,H]=[Y,D]=[Y,K]=0, \ \ \alpha, \beta = 1, 2,
\label{eq10}
\eea
\noindent along with the algebra of its $O(2,1)$ subgroup,
\be
[H,D] = i H, \ \ [H,K] = 2 i D, \ \ [D,K] = i K.
\label{eqo2}
\ee
\noindent The Casimir of the $O(2,1)$ is $C=\frac{1}{2} ( H K + K H ) - D^2$.
The quadratic and the cubic Casimir operators of
$OSp(2|2)$ are given by\cite{superlie,rev},
\bea
&& C_2= C + \frac{i}{4} [Q_1,S_1] + \frac{i}{4} [Q_2,S_2] - Y^2,\nonumber \\
&& C_3= C_2 Y - \frac{Y}{2} + \frac{i}{8} {\Big (} [Q_1,S_1] Y +
[Q_2,S_2] Y + [S_1,Q_2] D\nonumber \\
&& \ \ \ \ - [S_2,Q_1] D + [Q_1,Q_2] K + [S_1,S_2] H  {\Big )}.
\label{casi1}
\eea
\noindent The particular form of the cubic Casimir $C_3$ has been derived
from the general expression given in Ref. \cite{rev}. We will show that both
$C_2$ and $C_3$ are zero for the `atypical' superextension, while non-zero
for the `standard' superextension.

The subgroup $OSp(1|1)$ of $OSp(2|2)$ is described either by the set of
generators $A_1 \equiv \{ H, D, K, Q_1, S_1 \}$ or
$A_2 \equiv \{ H, D, K, Q_2, S_2 \}$. Let us first consider the
set $A_1$. The Casimir of the $OSp(1|1)$ is given by,
\be
C_1 = C + \frac{i}{4} [Q_1, S_1] + \frac{1}{16}.
\ee
\noindent One can define an even operator $C_s$,
\be
C_s = i [Q_1, S_1] - \frac{1}{2},
\ee
\noindent which has the property that it commutes with all the
bosonic generators and anticommutes with all the fermionic generators
of the set $A_1$.
Moreover, it satisfies, $C_1= \frac{1}{4} C_s^2$ and consequently,
$C=\frac{1}{4} C_s ( C_s-1) - \frac{3}{16}$. The operator $C_s$
is known as Scasimir and has important applications in the description
of the nonrelativistic dynamics of a spin-$\frac{1}{2}$ particle in the
background of monopoles, Dirac-Coulomb problem etc.\cite{dhowker,monopole}.
The Casimir and the Scasimir for the set $A_2$ are given by,
\be
\bar{C}_1 = C + \frac{i}{4} [Q_2, S_2] + \frac{1}{16}, \ \
\bar{C}_s = i [Q_2, S_2] - \frac{1}{2},
\ee
\noindent with the properties $\bar{C}_1=\frac{1}{4} \bar{C}_s^2$ and
$C=\frac{1}{4} \bar{C_s} (\bar{C_s}-1) - \frac{3}{16}$. This implies
that, in general, the Casimir $C$ can be factorized in two different ways,
either in terms of $C_s$ or $\bar{C}_s$.

We will show that the Scasimir $C_s$ and $\bar{C}_s$ are identical
for the `atypical' superextension. Consequently, the Casimir $C$ can be
factorized only in one way. Moreover, the Scasimir commutes with 
all the even operators $b$ and anticommutes with all the odd
operators $f$ of the $OSp(2|2)$. Thus, the Scasimir of the $OSp(1|1)$ can
be promoted to be the Scasimir of $OSp(2|2)$. On the other hand, for the
standard superextension, $C_s$ and $\bar{C}_s$ are different. Neither $C_s$
nor $\bar{C}_s$ or any linear combination of them can be promoted
to be the Scasimir of $OSp(2|2)$.

The generator of the compact rotation $R$ and the
creation-anhilation operator $B_{\pm}$ of the $O(2,1)$ are given by,
\be
R=\frac{1}{2} (H+K), \ \ B_{\pm} = \frac{1}{2} K - \frac{1}{2} H
\pm i D,
\label{comp}
\ee
\noindent and satisfy the algebra,
\be
[R, B_{\pm} ] = \pm B_{\pm}, \ \ [B_-, B_+] = 2 R.
\ee
\noindent The structure-equations (\ref{eq10}) can be written in the
Cartan basis where $R$ and $Y$ are simultaneously diagonal.
\noindent In the Cartan basis, define the operators,
\be
{\cal{H}}_+ = R + Y, \ \ {\cal{H}}_- = R - Y.
\label{fundu}
\ee
\noindent We will show that both ${\cal{H}}_+$ and ${\cal{H}}_-$ have
different co-ordinate representations corresponding to `atypical' and
`standard' superextension. ${\cal{H}}_+$ is related to the ${\cal{H}}_-$
and the vice versa through an automorphism of the $OSp(2|2)$ algebra.
So, it is suffice to consider either of them as the defining
super-Hamiltonian.

\section{Co-ordinate representation}

We will now show that it is possible to choose two different co-ordinate
representations of the generators of $OSp(2|2)$. Our choice is such
that the generators of the subgroup $OSp(1|1)$, namely $A_1 \equiv \{ H, D, K,
Q_1, S_1 \}$, have the same co-ordinate representation for both the
atypical and standard superextension. It is only the generators that
enlarge the $OSp(1|1)$ to $OSp(2|2)$, namely $\{Q_2, S_2, Y \}$, that are 
chosen to be different for these two cases. We follow the convention
of considering $(Q_2, S_2, Y)$ as part of the generators of the
`standard' superextension, while $(\hat{Q}_2, \hat{S}_2, \hat{Y})$ as that
of the `atypical' superextension. Consequently, any new operator/generator
that is defined using any one of the operators $(\hat{Q}_2, \hat{S}_2,
\hat{Y})$ will be denoted by $\hat{O}$ to distinguish it from $O$ that
is constructed out of $(Q_2, S_2, Y)$.

The operator $R$ is defined solely in terms of
$H$, $D$ and $K$ in Eq. (\ref{comp}). So, it has the same
co-ordinate representation for both the `standard' and the
`atypical' superextension. On the other hand, $Y=\frac{1}{2} \{Q_1, S_2 \}$
and $\hat{Y}=\frac{1}{2} \{Q_1, \hat{S}_2 \}$ are chosen to
be different for these two cases. It is now obvious from
Eq. (\ref{fundu}) that both the Hamiltonian ${\cal{H}}_{\pm}= R \pm Y$ and
$\hat{\cal{H}}_{\pm}= R \pm \hat{Y}$ have different co-ordinate representations,
corresponding to `atypical' and `standard' superextension.

We introduce $2 N$ entities $\xi_i$ satisfying the following
real Clifford algebra,
\be
\{\xi_i, \xi_j \} = 2 g_{ij}, \ \ g_{ij}=  \pmatrix{ {-I}& {0}\cr
{0} & {I}\cr},
\ee
\noindent where $I$ is an $N \times N$ identity matrix. The signature of the
metric $g_{ij}$ is such that the square of the $\xi_i$'s is equal to $1$ or
$-1$ depending on whether  $ i > N$ or $ i \leq N$, respectively. In
particular,
\be
\xi_{N+i}^2 = - \xi_{i}^2=1, \ \ i=1,2, \dots N.
\ee
\noindent It is known that one can introduce a new operator
$\gamma_5$\cite{clifford},
\be
\gamma_5 =  \xi_1 \xi_2 \dots \xi_{2 N-1} \xi_{2 N},
\ee
\noindent which anticommutes with all the $\xi_i$'s. Further, we can introduce
two idempotent operators $\gamma_5^{\pm}= \frac{1}{2} ( 1 \pm \gamma_5 )$
with $(\gamma_5^{\pm})^2 = \gamma_5^{\pm}$ and
$\gamma_5^+ \gamma_5^-=\gamma_5^- \gamma_5^+=0$,
since $\gamma_5^2 = 1$. The operators $\gamma_5$ and $\gamma_5^{\pm}$
play an important role in constructing independent real supercharges.

Most of our discussion will be based on the above real Clifford algebra.
However, we also need the complexified version of the above algebra
in order to make connection with the known results in the literature.
We define a set of fermionic variables $\psi_i$ and
their conjugates $\psi_i^{\dagger}$ as,
\be
\psi_i = \frac{i}{2} \left ( \xi_i - \xi_{N+i} \right ), \ \
\psi_i^{\dagger} = \frac{i}{2} \left ( \xi_i + \xi_{N+i} \right ).
\ee
\noindent These fermionic variables satisfy the Clifford algebra,
\be
\{\psi_i, \psi_j \} = \{\psi_i^{\dagger}, \psi_j^{\dagger} \} =0, \ \
\{\psi_i, \psi_j^{\dagger} \} = \delta_{ij}.
\ee
\noindent The fermionic vacuum $|0 \rangle$ and its conjugate $|\bar{0}
\rangle$ in the $2^N$ dimensional fermionic Fock space are
defined as, $\psi_i |0 \rangle = 0, \psi_i^{\dagger} |\bar{0} \rangle=0$.
The fermion number corresponding to each particle is defined
as, $n_i=\psi_i^{\dagger} \psi_i$ and the total fermion number
$n= \sum_i n_i$. The total fermion number $n=0$ for the fermionic vacuum and
$n=N$ for the conjugate vacuum. An equivalent expression for $\gamma_5$ in
terms of $n_i$ can be written as,
\be
\gamma_5 = (-1)^N \prod_{i=1}^N \left ( 2 n_i - 1 \right ). \ \
\ee
\noindent The action of $\gamma_5$ on a state $|n \rangle$ with fermion
number $n$ is, $\gamma_5 |n \rangle = (-1)^n |n \rangle$, where
$0 \leq n \leq N$. Note that $\gamma_5$ leaves the fermionic vacuum invariant,
i.e., $\gamma_5 |0 \rangle = |0 \rangle $. On the other hand,
$\gamma_5 |\bar{0} \rangle = (-1)^N |\bar{0} \rangle $, implying that the
conjugate vacuum is invariant for even $N$ and changes sign for odd $N$.

Consider the following Hamiltonian,
\be
H = \frac{1}{2} \sum_{i=1}^N  \left ( p_i^2 + W_i^2 \right )
-\frac{1}{2} \sum_{i,j=1}^N \xi_i \xi_{N+j} W_{ij}, \ \
W_i = \frac{\partial W}{\partial x_i}, \ \
W_{ij} = \frac{\partial^2 W}{\partial x_i \partial x_j}.
\label{hami}
\ee
\noindent The superpotential $W = ln \ G$ with $G$ being a
homogeneous function of degree $d$. This ensures that the bosonic potential
in $H$ scales inverse-squarely. Subsequently, we define the Dilatation operator
$D$ and the conformal operator $K$ as,
\be
D = - \frac{1}{4} \sum_{i=1}^N  \left ( x_i p_i + p_i x_i \right ), \ \
K = \frac{1}{2} \sum_i x_i^2.
\label{dami}
\ee
\noindent The operators $H$, $D$ and $K$ together satisfy an $O(2,1)$ algebra
(\ref{eqo2}).
We now introduce the supercharges $Q_1$ and $S_1$,
\be
Q_1 =  \frac{1}{\sqrt{2}} \sum_{i=1}^N \left [ - i \ \xi_i \ p_i +
\xi_{N+i} \ W_i \right ],\ \
S_1 = - \frac{i}{\sqrt{2}} \sum_{i=1}^N \xi_i \ x_i.
\label{q1s1}
\ee
\noindent The supercharges $Q_1$ and $S_1$ squares to the Hamiltonian $H$
and the conformal generator $K$, respectively.

The Casimir $C$ of the $O(2,1)$
algebra can be shown to be equivalent to the angular part of the $H$ in the
$N$-dimensional hyper-spherical co-ordinates, up to an additional
constant shift. In particular,
\be
C = \frac{1}{4} \left [ \sum_{i < j} L_{ij}^2 +
r^2 \left ( \sum_{i=1}^N W_i^2 -
\sum_{i,j=1}^N \xi_i \xi_{N+j} W_{ij} \right )
+ \frac{1}{4} N (N-4) \right ],
\label{casi}
\ee
\noindent where $L_{ij}= x_i p_j - x_j p_i$ and $r^2=\sum_i x_i^2$.
Note that both $W_i^2$ and $W_{ij}$ scale inverse-squarely. So, in
the $N$ dimensional hyperspherical co-ordinates $ r^2 ( \sum_i W_i^2
- \sum_{i,j} \xi_i \xi_{N+j} W_{ij} )$ contains only angular variables.
In Ref. \cite{gamb}, such a relation between the Casimir operator
and the angular part of a Hamiltonian with dynamical $O(2,1)$ symmetry
and involving only bosonic co-ordinates
was derived. Eq. (\ref{casi}) generalizes the work of \cite{gamb} by
including both bosonic and fermionic degrees of freedom. Note that
the relation between the Casimir and the angular part of the Hamiltonian
is valid as long as there is an underlying $O(2,1)$ symmetry and does not
depend on whether the Hamiltonian is supersymmetric or not.

\subsection{Standard superextension}

We consider the case of `standard' superextension in this subsection.
We first present our results in terms of real supercharges and at the end
of this subsection, we use complex supercharges to make correspondence
with the current literature. To the best of our knowledge, the many-particle
supersymmetric quantum mechanics with dynamical $OSp(2|2)$ supersymmetry has
never been discussed in terms of real supercharges. Although both the
formulation give the same result, the formulation in terms of real
supercharges brings out certain subtle features in connection with the
Casimir's of the model. We believe such features have never been discussed
previously.

The generators $Q_2$, $S_2$ and $Y$ are given by,
\bea
Q_2 & = & - \frac{1}{\sqrt{2}} \sum_i \left [ \xi_{N+i} \ p_i + i 
\xi_i W_i \right ],\nonumber \\
S_2 & = & - \frac{1}{\sqrt{2}} \sum_i \xi_{N+i} x_i,\nonumber \\
Y & = & - \frac{1}{4} \left ( \sum_i \xi_{N+i} \ \xi_{i} +  2 d \right ).
\label{q2s2}
\eea
\noindent The structure-equations (\ref{eq10}) of $OSp(2|2)$ are satisfied
with the above co-ordinate representation of the generators. In order to
see that the Casimir operators $C_2$ and $C_3$ are non-zero, we first find,
\bea
i [Q_1, S_1] & = & \frac{N}{2} + \frac{i}{2} \sum_{i,j=1}^N \xi_i \
\xi_j \ L_{ij} + \sum_{i,j=1}^N \xi_{N+i} \ \xi_j \ x_j \ W_i,\nonumber \\
i [Q_2, S_2] &  =  & \frac{N}{2} - \frac{i}{2} \sum_{i,j=1}^N \xi_{N+i} \
\xi_{N+j} \ L_{ij} - \sum_{i,j=1}^N \xi_{N+i} \ \xi_j \ x_i \ W_j,\nonumber \\
Y^2 & = & \frac{1}{16} \left ( 4 d^2 + N + 4 d \sum_i \xi_{N+i} \xi_i
-\frac{1}{2} \sum_{i \neq j} \left [ \xi_{N+i},
\xi_{N+j} \right ] \xi_i \xi_j \right ).
\label{wonda}
\eea
\noindent Note that $C$ in Eq. (\ref{casi}) contains a term proportional
to $ L_{ij}^2$. On the other hand, $i [Q_1,S_1] + i [Q_2,S_2]$
does not contain any term proportional to $L_{ij}^2$. It only contains a
term proportion to $L_{ij}$ and $Y^2$ is independent of bosonic coordinates.
This implies that $C_2$ in (\ref{casi1}) can not be zero.
Similarly, the first term of $C_3$ in (\ref{casi1}) contains a term of
the form $ L_{ij}^2 Y$. However, such a term can not be generated
from rest of the terms in the definition of $C_3$. So, the
cubic Casimir $C_3$ is also non-zero. It should be noted here that
although the Casimir's $C_2$ and $C_3$ are nonzero in terms of the coordinate
representation, their eigenvalues can be zero for the special case
corresponding to the atypical representation of the $OSp(2|2)$ algebra.

The coordinate representation of the operators $C_s=i [Q_1,S_1]-\frac{1}{2}$
and $\bar{C}_s = i [Q_2,S_2] -\frac{1}{2}$ can be simply found from the
first two equations of (\ref{wonda}). Recall that $C_s$ is the Scasimir of
the $OSp(1|1)$ corresponding to the set of generators $A_1$. Similarly,
$\bar{C}_s$ is the Scasimir of the $OSp(1|1)$ corresponding to the set of
generators $A_2$. We find that $C_s$ neither commutes
with $Y$ nor anticommutes with $Q_2$ and $S_2$. So, $C_s$ can not be considered
as the Scasimir of the full $OSp(2|2)$. Similarly, $[\bar{C}_s, Y] \neq 0$ and
$\{Q_1,\bar{C}_s \} \neq 0 \neq \{S_1,\bar{C}_S \}$. The operator $\bar{C}_s$
either can not be considered as the Scasimir of the $OSp(2|2)$. An explicit
calculation shows that $[C_s + \bar{C}_s, Y]=0$. However, the anticommutators
between $C_s  + \bar{C}_s$ and the supercharges $Q_1, Q_2, S_1, S_2$ are
nonzero. Thus, we conclude that the Scasimir's of the subgroup $OSp(1|1)$
or any linear combination of them can not be promoted to be the Scasimir
of the full $OSp(2|2)$.

A few comments are in order at this point. The Casimir $C$ of the $O(2,1)$ is 
equivalent to the angular part of the Hamiltonian $H$ in $N$ dimensional
hyper-spherical coordinates. We have seen that $C$ can be factorized either
in terms of $C_s$ or $\bar{C}_s$. Both $C_s$ and $\bar{C}_s$ commute with
$H$ and between themselves. So, $C_s$, $\bar{C}_s$ and $C$ can be diagonalized
simultaneously. Both $C_s$ and $\bar{C}_s$ being first order differential
operator, it may be much easier to solve them compared to $C$. Once the
eigenvalues of $C_s, \bar{C}_s$ are known, the eigenvalues of $C$ can be
computed easily. Secondly, the same method can also be used for a Hamiltonian,
\be
H^{\prime} = H + V(r),
\ee
\noindent for which the supersymmetry is explicitly broken by the introduction
of an arbitrary potential $V(r)$. Note that the angular part of $H^{\prime}$
and $H$ are still the same. Further, both $C_s$ and $\bar{C}_s$ commutes
with $H^{\prime}$. So, $C_s$ and $\bar{C}_s$ can be used to solve 
the non-supersymmetric Hamiltonian $H^{\prime}$! This is the first example of
${\cal{N}}=0$ supersymmetry\cite{spec} in the context of many-particle
quantum mechanics.

In order to see that the co-ordinate representation given by eqs.
(\ref{hami},\ref{dami},\ref{q1s1},\ref{casi},\ref{q2s2})
indeed corresponds to the standard superextension, we
write the super-Hamiltonian ${\cal{H}}_{\pm}$ in terms of $\psi_i$
and $\psi_i^{\dagger}$ as,
\be
{\cal{H}}_{\pm} =  \frac{1}{4} \sum_i \left ( p_i^2 + W_i^2 + x_i^2 \right )
+ \frac{1}{4} \sum_{i,j} \left [ \psi_i, \psi_j^{\dagger} \right ] W_{ij}
\pm \frac{1}{2} \left ( n - \frac{N}{2} - d \right ).
\label{h1}
\ee
\noindent
This form matches with the super-Hamiltonian with dynamical
$OSp(2|2)$ supersymmetry that was constructed in Ref. \cite{susy}. 
The supercharges are,
\be
F_{\pm}^L = \pm \frac{i}{2} \sum \psi_i \left ( p_i - i W_i \pm i x_i \right),
\ \ F_{\pm}^{R} =\pm \frac{i}{2} \sum \psi_i^{\dagger} \left
( p_i + i W_i \pm i x_i \right), \ \ \left ( F_{\pm}^L \right )^{\dagger}
= F_{\mp}^R,
\ee
\noindent with ${\cal{H}}_{\pm} = \{F_{\pm}^L, F_{\mp}^{R} \}$.
The total fermion number $n$ commutes with ${\cal{H}}_{\pm}$. Projecting
${\cal{H}}_+$ in the zero fermion sector and ${\cal{H}}_-$ in the
N-fermion sector, we get the purely bosonic Hamiltonian ${\cal{H}}_{\pm}^b$.
In particular,
\be
{\cal{H}}_{\pm}^b =  \frac{1}{4} \sum_i \left ( p_i^2 + W_i^2 \pm W_{ii}
+ x_i^2 \right ) 
- \frac{1}{2} \left (\frac{N}{2} \pm d \right ).
\label{hb}
\ee
\noindent The bosonic Hamiltonian ${\cal{H}}_{\pm}^b$ belongs to the general
class of Hamiltonian producing Calogero models\cite{susy}. The ground state
wavefunction of ${\cal{H}}_{+}^b$ is, $\phi_{+}=e^{ W - \frac{1}{2}
\sum_i x_i^2}$, while that of ${\cal{H}}_{-}^b$ is,
$\phi_{-}=e^{- W - \frac{1}{2} \sum_i x_i^2}$. The ground-state energy is
zero for both the cases. The parameters of ${\cal{H}}_{\pm}^b$ are contained
in the superpotential $W$. Note that $\phi_+^b$ and $\phi_-^b$ can not be
square integrable for the same range of parameters, since $W=ln \ G$ with
$G$ a homogeneous function. Thus, the admissible range of parameters due
to the square integrability of wave-functions, is different for
${\cal{H}}_{+}^b$ and ${\cal{H}}_{-}^b$
 
\subsection{Atypical Superextension}

We introduce the supercharges $\hat{Q}_2$, $\hat{S}_2$ and the bosonic
generator $\hat{Y} = \frac{1}{2} \{Q_1, \hat{S}_2 \}$ for the `atypical'
superextension as\cite{dhowker},
\bea
&& \hat{Q}_2 = -i \gamma_5 Q_1, \ \
\hat{S}_2= -i \gamma_5 \ S_1, \nonumber \\
&& \hat{Y}= \frac{\gamma_5}{2}
\left [ \frac{i}{2} \sum_{i,j} \xi_i \xi_j L_{ij} 
+ \sum_{i,j} \xi_{N+i} \xi_j W_i x_j
+ \frac{N}{2} \right ].
\label{eq9}
\eea
\noindent The above co-ordinate representation of the generators is 
not identical to the one given in Eq. (\ref{q2s2} ), only if the total number
of particles $N$ is greater than or at least equal to two, i.e. $N \geq 2$.
So, our results
emerge as the many-particle or higher dimensional features of the usual
one dimensional supersymmetric quantum mechanics.
The structure-equations (\ref{eq10}) are satisfied by the
bosonic generators $\{H, D, K, \hat{Y} \}$ and the fermionic
generators $\{Q_1, \hat{Q}_2, S_1, \hat{S}_2 \}$. The cubic Casimir now
takes a simple form \cite{dhowker}, $ C_3= (\hat{Y} -
\frac{\gamma_5}{4} ) C_2$. Using the defining
relation for $\hat{Y}$, we find $\hat{Y}= \frac{\gamma_5}{2} ( C_s +
\frac{1}{2})$ and consequently, $\hat{Y}^2 = \frac{1}{4} (C_s^2 + C_s +
\frac{1}{4})$. Expressing all other terms in $C_2$ in terms of $C_s$,
it is now easy to show that,
\be
C_2=0, \ \ C_3= (\hat{Y} - \frac{\gamma_5}{4} ) C_2 =0.
\ee
\noindent The spectrum is not completely specified by the eigenvalues of
the Casimir's and the representation is necessarily `atypical'. 
Further, note that the operators $C_s=\hat{\bar{C}}_s$.
Using the structure-equations of $OSp(2|2)$, we find that $C_s(\hat{\bar{C}}_s)$
commutes with $H$, $D$, $K$, $\hat{Y}$ and anti-commutes with $Q_1,
\hat{Q}_2, S_1, \hat{S}_2$. So, the operator $C_s$ is also the Scasimir
of the $OSp(2|2)$ supergroup and can be used to determine
the eigenspectrum.

As in the case of standard superextension, ${\cal{N}}=0$ supersymmetry is also
present for the `atypical' superextension. The only difference being that
we now have only one independent operator $C_s$ that factorized the Casimir
$C$. The co-ordinate representation of $C_s$ can still be determined from
the first equation of Eq. (\ref{wonda}) and it can be used to find the
spectrum of non-supersymmetric Hamiltonian $H^{\prime}$.

The super-Hamiltonian $\hat{\cal{H}}_{\pm}$ can be written in terms of
$\psi_i$ and $\psi_i^{\dagger}$ as,
\bea
\hat{\cal{H}}_{\pm} & = & \frac{1}{4} \sum_i \left ( p_i^2 + W_i^2 +
x_i^2 \right ) + \frac{1}{4} \sum_{i,j} \left [ \psi_i,
\psi_j^{\dagger} \right ] W_{ij}
\pm \frac{\gamma_5}{4} {\Big [}  N + 2 d \nonumber \\
& - & i \sum_{i,j} \left ( \psi_i^{\dagger} \psi_j^{\dagger} +
\psi_i^{\dagger} \psi_j \right ) e^W L_{ij} e^{-W} - i
\sum_{i,j} \left ( \psi_i \psi_j -
\psi_j^{\dagger} \psi_i  \right)  e^{-W} L_{ij} e^{W}\nonumber \\ 
& - & \sum_{i,j} \left ( \psi_i^{\dagger} \psi_j + \psi_j^{\dagger}
\psi_i \right ) \left ( x_i W_j + x_j W_i \right ) {\Big ]}.
\label{h2}
\eea
\noindent The supercharges are,
\be
\hat{F}_{\pm}^L = \gamma_5^{-} \left ( F_{\pm}^{R} + F_{\pm}^L \right ), \ \
\hat{F}_{\pm}^{R} = \gamma_5^{+} \left ( F_{\pm}^{R} + F_{\pm}^L \right ),\ \
\left ( \hat{F}_{\pm}^L \right )^{\dagger} = \hat{F}_{\mp}^R,
\ee
\noindent with $\hat{\cal{H}}_{\pm} = \{ \hat{F}_{\pm}^L,
\hat{F}_{\mp}^{R} \}$. Comparing (\ref{h1}) and (\ref{h2}), it is
obvious that ${\cal{H}}_{\pm}$ is different from $\hat{\cal{H}}_{\pm}$.
However, the total fermion number $n$ is not a conserved quantity for
$\hat{\cal{H}}_{\pm}$. Thus, the states can not be labeled in terms of
$n$. Note that the fermion-number violating terms appear only in the expression
for $\hat{Y}$. The operator $\hat{Y}$ is simultaneously diagonalized with $R$
and, in general, eigenvalue of $Y$ can be any complex number\cite{superlie,rev}.
If we choose the eigenvalue of $\hat{Y}$ to be
${\cal{Y}}^{+}=\frac{1}{4} ( N + 2 d)$
and project $R$ to the fermionic vacuum $|0 \rangle$, we see that
$\hat{\cal{H}}_{-}$ contains the purely bosonic Hamiltonian
${\cal{H}}_{+}^b$. Similarly, if we choose the eigenvalue of $\hat{Y}$ to be
${\cal{Y}}^{-}= - \frac{1}{4} ( N - 2 d)$
and project $R$ to the conjugate fermionic vacuum $|\bar{0} \rangle$, the purely
bosonic Hamiltonian ${\cal{H}}_{-}^b$ appears from $\hat{\cal{H}}_{+}$.
The eigenstates of $\hat{Y}$ corresponding to the eigenvalues
${\cal{Y}}^{\pm}$ are $\phi^+ |0 \rangle$ and $\phi^- |\bar{0} \rangle$,
respectively. Note that $\phi^+ |0 \rangle$ and $\phi^- |\bar{0} \rangle$ are
simultaneous eigenstates
of $R$. We conclude that the purely bosonic Hamiltonian ${\cal{H}}_{\pm}^b$
can be obtained from both ${\cal{H}}_{\pm}$ and$ \hat{\cal{H}}_{\pm}$ through
appropriate projections.

Finally, a few comments are in order. We have followed the convention of having
the same coordinate representation for the generators $H, D, K, Q_1, S_1$
for both standard and the atypical superextension. The co-ordinate
representation of $Q_2, S_2$ and $Y$ are different for these two cases. We
mention here about two other possibilities. Let us define,
\be
\hat{Q}_1=-i \gamma_5 Q_2, \ \ \hat{S}_1=-i \gamma_5 S_2, \ \
{\widehat{Y}} = \frac{1}{2} \{\hat{Q}_1, S_2 \}.
\ee
\noindent The operators $\{ \hat{Q}_1, Q_2, \hat{S}_1, S_2, H, D, K,
{\widehat{Y}} \}$ constitute a second `atypical' superextension of the
bosonic Hamiltonian ${\cal{H}}_{\pm}^b$. Note that
${\widehat{Y}}=-\frac{i \gamma_5}{2} [Q_2, S_2]$ and
$\hat{Y}=\frac{i \gamma_5}{2} [Q_1, S_1]$, where the commutators $[Q_1,S_1]$
and $[Q_2,S_2]$ are determined from Eq. (\ref{wonda}). So, the first chiral
super-model $\hat{\cal{H}}_{\pm}$ and the second chiral super-model
${\widehat{\cal{H}}} = R \pm {\widehat{Y}}$ have
different co-ordinate representations. Similarly,
$\{ \hat{Q}_1, \hat{Q}_2, \hat{S}_1, \hat{S}_2, H, D, K, -Y \}$ constitute
another `standard' superextension of the bosonic Hamiltonian
${\cal{H}}_{\pm}^b$. However, in this case, ${\cal{H}}_{\pm}$ of Eq.
(\ref{h1}) simply changes to ${\cal{H}}_{\mp}$.

One may ask at this point whether the supercharges $\{Q_1, Q_2, \hat{Q}_1,
\hat{Q}_2 \}$ can be used to construct an extended supersymmetry involving
the Hamiltonian $H$. Although each of these supercharges squares to $H$, 
some of the anti-commutators involving two different supercharges are non-zero.
In particular, we find the following relations,
\bea
&& H=Q_1^2=Q_2^2=\hat{Q}_1^2=\hat{Q}_2^2,\nonumber \\
&& \{Q_1, Q_2 \}=\{\hat{Q}_1, \hat{Q}_2\}=\{Q_1, \hat{Q}_2\}=
\{Q_2, \hat{Q}_1\}=0,\nonumber \\
&& \{Q_1, \hat{Q}_1 \}= i \gamma_5 [Q_1, Q_2], \ \
\{Q_2, \hat{Q}_2 \}= - i \gamma_5 [Q_1, Q_2],
\eea
\noindent where the last two anti-commutators are nonzero. So, an extended
supersymmetry for $H$ with the supercharges $\{Q_1, Q_2, \hat{Q}_1,
\hat{Q}_2 \}$ can not be constructed.

\section{Examples}

Our discussions so far have been confined to the general
superpotential $W$. Specific choices of $W$ lead to different interesting
physical models. We consider here three examples, (i) free superoscillators,
(ii) superconformal quantum mechanics and (iii) super-Calogero models. The
`standard' super-models for these three cases are exactly solvable.
Although the representation of the $OSp(2|2)$ is necessarily `atypical' for
chiral super-models, the spectrum can be obtained exactly following
\cite{dhowker}. However, the spectrum generating algebra may be bigger than
$OSp(2|2)$ for some cases. This is definitely the case for chiral
super-Calogero model. The study of the complete spectrum
of the chiral super-models thus needs a detail investigation, which is beyond
the scope of this paper. We write down below only the standard and the chiral
super-Hamiltonian and their ground states for the three examples considered in
this paper. The excited states belonging to the $OSp(2|2)$ multiplet may be
obtained\cite{dhowker} by acting the strings of creation operators
$B_+, \hat{F}_{+}^L, \hat{F}_{+}^R$ on the groundstate. Note that
$\hat{F}_{+}^{L,R}$ is linear in both $\psi_i$ and $\psi_i^{\dagger}$, which
is consistent with the fact that fermion number $n$ is not a conserved quantity
for the chiral super-models.

\subsection{Superoscillators}

Consider the superpotential $W=0$. Consequently, we have to take $d=0$,
which is the degree of homogeneity of $G$.
The Hamiltonian $H$ is that of $N$ free superparticles. Both
${\cal{H}}_{\pm}$ and $\hat{\cal{H}}_{\pm}$
describe the Hamiltonian of $N$ free superoscillators in different co-ordinate
representation. In particular,
\bea
{\cal{H}}_{\pm} & = & \frac{1}{4} \sum_i \left ( p_i^2 + x_i^2 \right )
\pm \frac{1}{2} \left ( n - \frac{N}{2} \right ),\nonumber \\
\hat{\cal{H}}_{\pm} & = & \frac{1}{4} \sum_i \left ( p_i^2 +
x_i^2 \right ) \pm \frac{\gamma_5}{4} \left [ N - i \sum_{i,j}
\left ( \psi_i^{\dagger} \psi_j^{\dagger} + \psi_i \psi_j  +
\psi_i^{\dagger} \psi_j - \psi_j^{\dagger} \psi_i
\right) L_{ij} \right ].
\eea
\noindent Note that for $N=1$, ${\cal{H}}_{\pm}$ and $\hat{\cal{H}}_{\mp}$
are identical. The differences show up only for $N \geq 2$.
The groundstates of ${\cal{H}}_{\pm}$ are $e^{-\frac{1}{2} r^2} |0 \rangle$
and $e^{-\frac{1}{2} r^2} |\bar{0} \rangle$, respectively. The ground-state
energy
is zero for both the cases. On the other hand, the zero-energy groundstates of
$\hat{\cal{H}}_{\pm}$ are $e^{-\frac{1}{2} r^2} |odd \rangle$ and
$e^{-\frac{1}{2} r^2} |even \rangle$, respectively. The symbol $|odd \rangle
( |even \rangle )$ denotes a fermionic state with odd(even) fermion number $n$.
It is a very special feature of the chiral superoscillators that the
groundstate can be constructed
using any odd (even) fermion number $n$. For $W \neq 0$, it turns out that the
even state is necessarily the fermionic vacuum $|0 \rangle$ and the odd state
is the conjugate vacuum $|\bar{0} \rangle$ for odd $N$. The total fermion
number is not a conserved quantity for $\hat{\cal{H}}_{\pm}$. So, 
linear combinations of states with odd(even) fermion numbers are also
exact eigen states of superoscillators. However, linear combination of states
with with odd and even fermion numbers are not valid eigenstates, since mixing
of different chiralities are not allowed.

\subsection{Super-conformal quantum mechanics}

We choose $W= \eta \ ln r$ and consequently, $d=\eta$.
The Hamiltonian $H$ for this case is an example of superconformal quantum
mechanics. The Hamiltonian $H$ has no normalizable ground-states and
the prescription\cite{fr} is to study the evolution of the operator
${\cal{H}}_{\pm}$. The new representation suggests that an alternative
prescription may be to study the evolution of the chiral Hamiltonian
$\hat{\cal{H}}_{\pm}$ instead of ${\cal{H}}_{\pm}$.
In particular, these Hamiltonians are given by,
\bea
{\cal{H}}_{\pm} & = & \frac{1}{4} \left ( \sum_i p_i^2 +
\frac{\eta(\eta-N+2)}{r^2} + r^2 \right )- 
\frac{\eta}{2 r^2} \sum_{i,j} \psi_j^{\dagger} \psi_i \left (\delta_{ij}
- 2 \frac{x_i x_j}{r^2} \right )
\pm \frac{1}{2} \left ( n - \frac{N}{2} - \eta \right ),\nonumber \\
\hat{\cal{H}}_{\pm} & = & \frac{1}{4} \left ( \sum_i p_i^2 +
\frac{\eta(\eta-N+2)}{r^2} + 
r^2 \right ) - \frac{\eta}{2 r^2} \sum_{i,j} 
\psi_j^{\dagger} \psi_i \left (\delta_{ij} - 2 \frac{x_i x_j}{r^2} \right ) 
\pm \frac{\gamma_5}{4} {\Big [} N + 2 \eta\nonumber \\ 
& - & i \sum_{i,j} \left ( \psi_i^{\dagger} \psi_j^{\dagger} +
\psi_i^{\dagger} \psi_j + \psi_i \psi_j -
\psi_j^{\dagger} \psi_i  \right) L_{ij} 
-\frac{2 \eta }{r^2} \sum_{i,j} \left ( \psi_i^{\dagger} \psi_j +
\psi_j^{\dagger} \psi_i \right ) x_i x_j {\Big ]}.
\label{ho}
\eea
\noindent Note that for $N=1$, ${\cal{H}}_{\pm}$ and $\hat{\cal{H}}_{\mp}$
are identical. The differences show up only for $N \geq 2$. For $\eta=0$,
the Hamiltonian of superoscillators is reproduced. The zero-energy groundstate
of ${\cal{H}}_{+}$ and $\hat{\cal{H}}_{-}$ is,
$r^{\eta} e^{-\frac{1}{2} r^2} |0 \rangle, \eta \geq 0$.
The restriction on $\eta$ comes from the self-adjointness of the
super-Hamiltonian.The zero-energy groundstate of ${\cal{H}}_{-}$
is $\chi=r^{-\eta} e^{-\frac{1}{2} r^2} |\bar{0} \rangle, \eta < 0$.
The same
wave-function $\chi$ is the zero-energy ground-state of $\hat{\cal{H}}_{+}$
for odd $N$ only. For even $N$, $\chi$ is an exact eigenstate of
$\hat{\cal{H}}_{+}$ with non-zero energy.
 
\subsection{Calogero Models}

The superpotential for the $A_{N+1}$-type supersymmetric rational
Calogero model is given by,
\be
G_{A_{N+1}} = \prod_{i<j} (x_i - x_j)^{g}, \ \ d=\frac{g}{2} N (N-1).
\ee
\noindent The Hamiltonian ${\cal{H}}_{\pm}$ and $\hat{\cal{H}}_{\pm}$ are,
\bea
{\cal{H}}_{\pm} & = & \frac{1}{4} \sum_i \left ( p_i^2 + x_i^2 +
\sum_{j \neq i} \frac{g(g-1)}{(x_i - x_j )^2} \right )
+ \frac{g}{2} \sum_{i\neq j} \frac{\psi_i^{\dagger} \psi_i -
\psi_j^{\dagger} \psi_i}{( x_i - x_j )^2}
\pm \frac{1}{2} {\Big (} n - \frac{N}{2} -
\frac{g}{2} N (N-1) {\Big )},\nonumber \\
\hat{\cal{H}}_{\pm} & = & \frac{1}{4} \sum_i \left ( p_i^2 + x_i^2 +
\sum_{j \neq i} \frac{g(g-1)}{(x_i - x_j )^2} \right )
+ \frac{g}{2} \sum_{i\neq j} \frac{\psi_i^{\dagger} \psi_i -
\psi_j^{\dagger} \psi_i}{( x_i - x_j )^2}
\pm \frac{\gamma_5}{4} {\Big [}  N + g N (N-1)\nonumber \\
& - & i \sum_{i,j} \left ( \psi_i^{\dagger} \psi_j^{\dagger} +
\psi_i^{\dagger} \psi_j \right ) \left ( G_{A_{N+1}} L_{ij} \
G_{A_{N+1}}^{-1} \right ) - i
\sum_{i,j} \left ( \psi_i \psi_j - \psi_j^{\dagger} \psi_i  \right)
\left ( G_{A_{N+1}}^{-1} L_{ij} \ G_{A_{N+1}} \right )\nonumber \\ 
& - & g \sum_{i,j} \left ( \psi_i^{\dagger} \psi_j + \psi_j^{\dagger}
\psi_i \right ) \left ( \sum_{k\neq j} x_i (x_j - x_k)^{-1}
+ \sum_{k \neq i} x_j (x_i - x_k )^{-1} \right ) {\Big ]}.
\eea
\noindent After separating out the centre of mass, the standard
super-Calogero model for $N=2$ is shown to be identical to the Hamiltonian
of $N=1$ superconformal quantum mechanics of Ref. \cite{fr}. For the chiral
super-Calogero model with $N=2$, the centre of mass can not be separated out
from the residual degree of freedom. In the limit of vanishing $g$ and
arbitrary $N$, we recover the case of free superoscillators.
The zero-energy groundstate of ${\cal{H}}_+$ and $\hat{\cal{H}}_-$
is $G \ e^{-\frac{1}{2} r^2} \ |0 \rangle, g > 0$. Similarly, the zero-energy
groundstate of ${\cal{H}}_-$ is $\chi_1=G^{-1} \ e^{-\frac{1}{2} r^2}
\ |\bar{0} \rangle, g < 0$. As in the case of superconformal quantum mechanics,
$\chi_1$ is the zero energy eigenstate of $\hat{\cal{H}}_+$ for odd $N$ only.
For even $N$, $\chi_1$ is an exact eigenstate of $\hat{\cal{H}}_+$ with nonzero
energy.

The superpotential for the $BC_{N}$-type supersymmetric rational
Calogero model is given by,
\be
G_{BC_N} (\lambda, \lambda_1, \lambda_2) = \prod_{i <j} \left ( x_i^2 -
x_j^2 \right )^{\lambda}
\prod_k x_k^{\lambda_1}  \prod_l (2 x_l)^{ \lambda_2}, \ \
d=\lambda N (N-1) + N \lambda_1 + N \lambda_2,
\label{eq12}
\ee
\noindent where $\lambda$, $\lambda_1$ and $\lambda_2$ are arbitrary
parameters. The $D_{N}$-type model is described by
$\lambda_1=\lambda_2=0$, while $\lambda_1=0 (\lambda_2=0)$ describes
$C_{N+1} (B_{N+1})$-type Hamiltonian. Any particular result for $BC_N$
type model can be obtained using this form of superpotential from the
equations in Sec. III, which we do not repeat here.

\section{Summary \& Discussions}

We have constructed the supersymmetric extension of the rational Calogero
model with $OSp(2|2)$ dynamical supersymmetry that is different from the one
in Ref. \cite{fm} in many respects. First of all, the super-Hamiltonian have
different co-ordinate representations in these two constructions. In the
new construction, the quadratic and the cubic Casimir operators are
necessarily zero, thereby having only `atypical' representation. Further,
the Scasimir of the subgroup $OSp(1|1)$ can be promoted to be the Scasimir
of the full $OSp(2|2)$. On the other hand, the cubic and the quadratic
Casimir's are, in general, non-zero for the standard construction. The
Scasimir of the $OSp(1|1)$ can not be promoted to be the Scasimir of the
full $OSp(2|2)$ in this case.

We have also shown that $C$, which is equivalent to the
angular part of the super-Hamiltonian $H$ in the N-dimensional
hyper-spherical coordinate, can be factorized in terms of first order
differential operators that commute with the Hamiltonian. Thus, eigen
spectrum of these first order operators can be used to find the eigenspectrum
of $C$. The result is valid for both standard and the atypical superextension.
Surprisingly, this was never realized before for the standard superextension.
The reformulation of the problem in terms of two real supercharges made us
to see this subtle connection. We have also shown that super-Calogero model,
for both the standard and the atypical superextension, can be deformed suitably
to admit ${\cal{N}}=0$ supersymmetry.

The supercharge $-\sqrt{2} Q_2$ can be identified with a $2 N$ dimensional
Euclidean Dirac operator which is independent of the bosonic coordinates
$x_{N+1}, x_{N+2}, \dots, x_{2 N}$. The supercharge $Q_2$ being part of the
$OSp(2|2)$ algebra, its spectrum can be readily derived algebraically.
For the special case of rational super-Calogero model of $A_{N+1}$-type,
\be
W_i = g \sum_{j(\neq i)} \frac{1}{x_i - x_j}, \ i, j=1, 2, \dots, N,
\ee
\noindent which describes a many-particle Coulomb interaction. The
mathematical as well as physical significance of this many-particle
Dirac-Coulomb operator needs to be well understood.

The standard super-Calogero model is exactly solvable, although the
$OSp(2|2)$ symmetry alone can not determine the
complete spectrum. The spectrum generating algebra is higher than the
$OSp(2|2)$. One would like to know at this point whether the 
chiral super-Calogero model is also exactly solvable
or not? Part of the spectrum corresponding to the $OSp(2|2)$ symmetry
can of course be obtained analytically. However, it is desirable to study
different aspects related to integrability and exact solvability
of the chiral super-Calogero model.

\acknowledgments{ I would like to thank Ryu Sasaki for useful comments on the
manuscript. This work is supported (DO No. SR/FTP/PS-06/2001) by SERC,
DST, Govt. of India through the Fast Track Scheme for Young
Scientists:2001-2002.}

\end{document}